**Thickness monitoring of graphene on SiC using low-energy electron diffraction**

P. J. Fisher
IBM T. J. Watson Research Center, Yorktown Heights, NY 10598

Luxmi, N. Srivastava, S. Nie,[*] and R. M. Feenstra[†]
Dept. Physics, Carnegie Mellon University, Pittsburgh, PA 15213

**Abstract**
The formation of epitaxial graphene on SiC is monitored *in-situ* using low-energy electron diffraction (LEED). The possibility of using LEED as an *in-situ* thickness monitor of the graphene is examined. The ratio of primary diffraction spot intensities for graphene compared to SiC is measured for a series of samples of known graphene thickness (determined using low-energy electron microscopy). It is found that this ratio is effective for determining graphene thicknesses in the range 1 to 3 monolayers. Effects of a *distribution* of graphene thicknesses on this method of thickness determination are considered.

[*] present address: Sandia National Laboratory, Livermore, CA
[†] feenstra@cmu.edu

## I. Introduction

Graphene, which refers to one or more monolayer sheets of sp$^2$ hybridized carbon, has been intensively studied for the past five years because of its unique electronic structure.[1] The formation of epitaxial graphene on SiC has developed over the past few years into an active subject of research.[2] The high mobility of the graphene holds promise for applications in high-speed electronics.[3] Although detailed measurements of mobility as a function of graphene thickness have not yet been made, it is reasonable to suppose that the highest mobilities will be realized for uniform graphene coverages of discrete, integer numbers of monolayers. To realize this potential, however, it is necessary to have a reproducible means of achieving certain specific coverages of graphene on the SiC.

The epitaxial graphene is formed by annealing the SiC at high temperatures for some amount of time, in vacuum or in some other controlled environment.[2,4,5] In principle, careful control of these parameters might be sufficient to enable the production of a specific thickness of graphene. In practice, however, variation in thermal contact between sample and heater produces significant uncertainty in the annealing conditions (and, in addition, optical pyrometry is problematic to use for temperature measurement since the SiC is transparent). Thus, it is desirable to have some *in-situ* means of graphene thickness measurement, to allow the monitoring of the graphene during its formation. Such a method would also be quite useful for studying new graphene formation environments. Low-energy electron microscopy (LEEM) seems to be the best available method for this purpose,[6] but the expense and complexity of this technique renders it inapplicable for routine graphene production systems. An alternative is to use low-energy electron diffraction (LEED), and indeed, a prior report by Riedl *et al.*[7] proposes using *intensity vs. energy spectra* of the primary (1,0) graphene LEED spot for thickness control. This method is also somewhat complex however, and as a simpler alternative those authors considered use of an *intensity ratio* (at 126 eV) of this graphene LEED spot compared to that of its neighboring 6√3×6√3-R30° satellite spots. This same intensity ratio (at 53.3 eV) was also considered by Virojanadara *et al.*,[5] although in both of these cases a *quantitative* relationship between the intensity ratio and the graphene thickness was not established.

In this work we establish a quantitative relationship between a LEED intensity ratio and the graphene thickness. Rather than using the 6√3 satellite spots mentioned above, we employ the ratio of the primary graphene spot to the primary SiC spot which we argue provides a somewhat more general method. A series of samples are prepared by vacuum annealing in a graphene production system (suitable for graphene preparation on semi-insulating SiC, and scalable to large wafer sizes) and monitored *in-situ* by LEED. LEEM is then used *ex-situ* to determine the graphene thickness, and a relationship between the LEED ratio and the thickness is established. Compared with other methods for *in-situ* thickness monitoring, we argue that this LEED intensity ratio provides a robust means of monitoring the graphene thickness even for situations where a distribution of graphene thicknesses and/or some surface disorder may be present.

## II. Experimental

Two different ultra-high-vacuum (UHV) systems were used to produce the graphene films studied in this work. The first was a standard surface science system equipped with a manipulator and VG LEED apparatus as well as a scanning tunneling microscope, as described elsewhere.[8] Heating of the SiC was accomplished in this system by resistive heating of conductive substrates, and monitoring of the LEED patterns was performed primarily in the regime *prior* to graphene formation and also up to 1 – 2 monolayers of graphene (graphene monolayer = ML = 38.0 carbon atoms/nm$^2$).

The second UHV system is dedicated to graphene production, containing a graphite strip heater and pumped by a 150 l/s turbo-molecular pump and a hydrogen-getter pump. A graphite plate with thickness 1 mm and area 100×75 mm$^2$ is cut into a bow-tie shape, with a narrow neck of 20 mm length and 14 mm width. Two thick (dual, 9.5 mm diameter) water-cooled copper feedthroughs are used to transmit the current, mounted onto large copper clamps on the two 75 mm ends of the plate. Current is supplied by a transformer capable of supplying up to 300 A at 6.3 V. A gate valve separates the turbo pump from the UHV chamber; the gate valve is closed for H-etching and it is open for the graphitization. A VG LEED system was installed in a geometry such that LEED data can be acquired from a sample resting on the heater. A Ta foil mounted on a manipulator is located about 5 mm above the sample. This foil is moved over to cover the sample during annealing, thereby protecting the LEED apparatus that is located directly above the sample. In addition, the foil conveniently acts as a reflector, trapping some of the blackbody radiation produced by the heater strip. In the *absence* of the Ta strip we have previously estimated a temperature difference between sample and heater of ≈400°C.[4] By examination of the graphene thickness formed on samples *with* the Ta reflector in place, we roughly estimate a temperature difference between sample and strip of ≈200°C for this configuration. Digital acquisition of the LEED patterns was accomplished using a video system and supporting software from k-Space Associates.

The experiments have been performed on nominally on-axis, conducting 6H-SiC substrates that were purchased from Cree Corp. As received, these substrates had been mechanically polished on the (0001) surface (the so-called Si-face) and they were "epi-ready" (*i.e.* with further polishing and a damage removal step). Samples measuring 10×10 mm$^2$ were cut from the wafers. After introduction in the UHV system H-etching was performed at 1 atm pressure, using 99.9995% purity hydrogen with a flow rate of 10 lpm and at a temperature of 1550°C for 3 min, to eliminate scratches. After pumping out the H to a pressure below 3×10$^{-8}$ Torr, graphitization was performed. The material used to fabricate the graphite heater strip was obtained from Poco Graphite, and is Semiconductor Grade material. No measurable contamination as seen by residual gas analysis is found to be emitted during the graphitization (these measurements were performed only after the first few heating runs with the strip). Low-energy electron microscopy was performed using an Elmitec LEEM III instrument.

### III. Results and Discussion
### A. Graphene Thickness Determination
For an absolute thickness determination of our graphene films we use LEEM. A typical LEEM image, acquired at 3.7 eV incident electron energy from a graphene film on Si-face SiC, is shown in Fig. 1(a). The results are very similar to those previously discussed by Hibino *et al.*, with oscillations in the emitted intensity of the electrons arising from interference between those electrons and the thickness of the graphene film,[6] as shown in Fig. 1(b). According to their analysis, the number of minima seen in the intensity curves between 0 and 10 eV corresponds to the number of graphene monolayers on the surface.

Thus, for regions A – C in Fig. 1, there are, respectively, 1, 2, and 3 ML of graphene on the surface. The majority of this surface is covered with 2 ML, with this coverage appearing as the mottled bright contrast in the image.[9] Video arrays of such images are acquired, extending over electron energies of 0 – 10 eV with spacing of 0.1 eV. These arrays of images are analyzed pixel by pixel, examining the oscillations in the intensity over the energy range of typically 2.0 to 6.5 eV. A quadratic background subtraction is performed on such curves, and the result is fit to a sine function with variable amplitude, period and phase. By viewing a scatterplot of the measured periods and phases, the ranges of period and phase corresponding to integer ML thicknesses can be identified. Hence the fraction of surface area covered by each integer thickness is determined. For example, in the image of Fig. 1 the fraction of surface covered by 1, 2, or 3 ML is 0.15, 0.81, and 0.04 respectively. Averaging over several such images acquired from different locations on the surface, we find an average graphene thickness for this sample of 1.92±0.08 ML.

**B. *In-situ* LEED**

Figures 2 and 3 show LEED results from Si-face SiC, acquired at room temperature after annealing the surfaces at consecutive steps for the times and temperatures specified. The data of Fig. 2 was acquired using our surface science system and is mainly concerned with the surface reconstructions occurring before and during the graphene formation; these results are in good agreement with those of prior authors.[10,11,12,13,14,15] The initial unreconstructed 1×1 pattern of the substrate is replaced by a $\sqrt{3}\times\sqrt{3}$-R30° reconstruction on heating to 900°C, the latter associated with one Si adatom per $\sqrt{3}\times\sqrt{3}$ SiC unit cell.[16] On further heating to 1200°C this pattern then progresses to a $6\sqrt{3}\times6\sqrt{3}$-R30° arrangement with a six-fold arrangement of spots around all of the $\sqrt{3}$ and 1×1 SiC spots. This $6\sqrt{3}$ arrangement is known to be associated with the C-rich reconstruction that first forms on the surface of the SiC and then persists at the graphene/SiC interface.[11,12,13,14,15,17,18,19] This so-called "buffer layer" contains nearly a full graphene monolayer's worth of excess carbon.[20] At 1250°C, the graphene (1,0) primary spots first appear and concomitantly the $\sqrt{3}$ and $2\sqrt{3}$ spots decrease in intensity [Fig. 2(d)]. Note in particular that the $\sqrt{3}$ spots decrease in intensity compared to their neighboring $6\sqrt{3}$ satellite spots, which is a characteristic feature of the onset of graphene formation.[8] Upon further annealing the intensity of the graphene pattern increases while that of the SiC decreases. A six-fold arrangement of $6\sqrt{3}$ satellite spots can be seen around the primary graphene spots in Fig. 2(h), acquired at an energy of 70 eV which enhances the intensity of those spots.

Figure 3 shows data from our graphene production system. The final sample configuration corresponding to Fig. 3(h) had a graphene thickness of 1.2 ML. The initial pattern is similar to that observed in Fig. 2(a), a simple SiC 1×1. After the hydrogen etch, however, a $\sqrt{3}\times\sqrt{3}$ pattern is observed. To our knowledge this is the first report of an *in-situ* surface structure following H-etching; this structure could be the Si-adatom arrangement mentioned above associated with Fig. 2(b), or perhaps it could be some other arrangement of the same symmetry. As a function of the annealing temperature we can clearly see a monotonic increase in the graphene primary spot intensity in the images from 1240°C (no visible graphene intensity) to 1400°C (where the intensity due to the graphene has finally surpassed that due to the underlying SiC). There are no visible satellite spots around the primary graphene spots in Figs. 3 because of the 100 eV energy used, but at an energy of 133 eV, shown in the inset of Fig. 3(h), 6√3 satellite spots are clearly observed in agreement with prior work.[7]

To determine the graphene thickness using LEED, we consider the intensity ratio of the primary graphene spots to the primary SiC spots. We choose this ratio, rather than using the 6√3 satellite spots employed by previous authors,[5,7] since we feel that such a technique might be more generally applicable for other situations, *e.g.* on the ($000\bar{1}$) surface (C-face) which does not display satellite spots.[21] We choose an energy of 100 eV, for which both of these primary spots are relatively intense. Within each set of spots the intensity varies somewhat as we go around the sixfold pattern, but this variation is very systematic from run to run and presumably arises from instrumental irregularities (*i.e.* variations in phosphor efficiency of the LEED screen and/or more magnetic field interference on some spots than others). We determine peak spot intensities by taking the average of the maximum pixel value together with the values of the four surrounding pixels, with a standard deviation computed from the same set. The results are then averaged over the six spots in the pattern (or five, if one is blocked by the mounting of the electron-gun). A background value is determined for the pattern by considering points located between the graphene and SiC spots. The resulting ratios are plotted in Fig. 4, as a function of the measured graphene thickness. The error on the latter is based on multiple measurements across each sample. For the data points in Fig. 4 with thickness greater than 1.5 ML we used LEEM to determine the thickness, but for the data point near 1.2 ML this sample was inadvertently damaged prior to LEEM study so we used its thickness as determined by Auger electron spectroscopy (AES), with the AES measurements carefully calibrated over many samples to LEEM data. We see in Fig. 4 that the LEED intensity ratio increases with the graphene thickness in a superlinear manner, as expected since the SiC spot intensity is decreasing as the graphene spot intensity increases.

Constructing a rigorous model for describing the LEED intensity ratio is, of course, very complicated because of the multiscattering that occurs in LEED. We therefore consider an approximate model in which there are layer-by-layer contributions to the LEED pattern. For example, for a single graphene layer we take the graphene spot intensity to be $I_1$ and the SiC intensity to be $I_0 \exp(-2t/\lambda)$ where $t$ is the graphene thickness, $\lambda$ is an attenuation length, and the factor of 2 is for attenuation of both the incident and diffracted electrons. If we have a fractional coverage $f \leq 1$ of the graphene, the graphene intensity would be $fI_1$ and the SiC intensity would be $I_0[(1-f)+f\beta]$ where $\beta \equiv \exp(-2t/\lambda)$. For a fractional coverage $1 \leq (1+f) \leq 2$ the graphene intensity is $I_1[1+f\beta]$ and SiC intensity is $I_0[(1-f)\beta + f\beta^2]$. For a fractional coverage $2 \leq (2+f) \leq 3$ the graphene intensity is $I_1[1+\beta+f\beta^2]$ and SiC intensity is $I_0[(1-f)\beta^2 + f\beta^3]$, and similarly for other coverages. Taking a ratio of these predicted intensities, we arrive at the solid line shown in Fig. 4, which is matched to the data for parameter values of $\beta = 0.29$ and $(I_1/I_0) = 0.24$.

In the above formulation of our model we neglected the effect of the 6√3×6√3-R30° arrangement that exists below the graphene. Various models for this so-called "buffer layer" exist, ranging from a distorted graphene layer to a C-rich reconstruction with much different structure.[22,23] In the former case some contribution of the buffer layer to the intensity of the graphene LEED spot is expected, thereby increasing that intensity and producing a theoretical ratio of graphene to SiC spot intensities that is *larger* than the solid line of Fig. 4, particularly for small graphene thicknesses. We can estimate this effect from the data of Fig. 2(d), which was acquired from a sample containing a mixture of 6√3×6√3-R30° and 5×5 surface structures (about 50% of each) but with essentially no graphene on top of the 6√3×6√3-R30° regions.[8] We find from that data a ratio of graphene to SiC spot intensities of ≈0.03, which, correcting for the 50% coverage of the buffer layer, yields a ratio of 0.06. Compared to the above $(I_1/I_0)$ value of 0.24 we find that the buffer layer contributes only about 1/4 as much to the graphene spot intensity as a graphene layer itself. Reworking our model to include this effect yields the dot-dashed line in Fig. 4, computed for the same parameters as the solid line. The effect of the buffer layer is thus seen to be rather small (although it should be noted that the $I_0$ value in the preceding discussion now corresponds to the intensity of the SiC spot as seen *through* the buffer layer).

Having established the curve shown in Fig. 4, we can use it to produce a desired graphene thickness. The observations are not real-time, since we must cool down the sample (and remove the protective Ta shield) prior to LEED observation, but nevertheless it can easily be accomplished in a matter of minutes. In practice, we initially anneal the sample for 5 min at some temperature (based on the heater current) slightly below that which we estimate is needed to achieve a given thickness and we then check the LEED intensity ratio. Depending on those results, further annealing steps, possibly at higher current, are used. With 3 or 4 such steps we can typically achieve a coverage within 10% of the desired one, and additional steps would permit finer control. In contrast, if we attempt to control the thickness based solely on the time and temperature, we achieve control of only about 30% at best, due to varying thermal contact between sample and heater. By internally calibrating a growth rate in this manner, for each run, the operator should be able to achieve custom/targeted thicknesses with an accuracy and consistency that processing techniques based on post-growth characterization feedback looping cannot reasonably obtain.

It is important to note that use of LEED for this application is limited to graphene thicknesses less than some saturation thickness (dependent on the beam energy), beyond which negligible intensity occurs from the SiC substrate. As seen in Fig. 4, this limiting thickness is slightly above 3 ML for the 100 eV energy that we use. Other techniques such as AES or Raman spectroscopy may not encounter this limitation due to their larger penetration depths. However, these other methods (AES in particular) are much less specific to the particular bonding arrangement of the carbon on the surface. The LEED intensity ratio used in this work is sensitive only to the carbon in the graphene, whereas AES would reveal the total carbon content. If, *e.g.*, some nano-crystalline graphite were to occur on the surface (as occurs in certain cases, especially for the $(000\bar{1})$ surface),[24] then AES would detect *all* of the carbon whereas LEEM would see only the carbon in the well-ordered graphene. The high surface sensitivity of LEED makes it ideal for studying graphene films with thickness up to a few monolayers, which are expected to be most useful for electronic applications. Of course, it might be useful to have *in-situ* measurements using several methods, to yield complementary information.

Another issue has to do with the *distribution* in thickness of the graphene. We can account for this effect by modification of the above model. We consider *e.g.* a Gaussian distribution of thickness, with the probability of having an *n*-monolayer region being proportional to $\exp[-(n-n')/(2\sigma^2)]$ where $n'$ is the most probable thickness and $\sigma$ is the width of the distribution. These probabilities are summed for $n = 0,1,2,3...$ ML and normalized, and the mean thickness $\bar{n}$ for the distribution is also computed. Using these probabilities together with results for integer coverages given above, we compute intensity ratios for this model. For the sample of Fig. 1, the standard deviation in the measured thickness from image to image was 0.08 ML, but evaluating instead the standard deviation within a *single* image we find 0.42 ML for Fig. 1(a), or 0.39 ML when averaged over additional images. In Fig. 4 we show by the dotted line results for a distribution width of $\sigma = 0.4$ ML. We see that a small shift in the results occurs compared to the layer by layer growth. Thus, it must be realized that this method of thickness calibration is somewhat dependent on the *distribution* of thicknesses (the same is true for the other methods as well).

**IV. Conclusions**
The progression of the surface structure of SiC upon annealing up to and above the graphitization temperature was studied *in-situ* by LEED, and results were compared to *ex-situ* analysis by LEEM. It was shown that the film/substrate (graphene/SiC) LEED intensity ratios, at 100 eV, track closely with the thickness of the growing graphene film. The model used here for analysis of LEED intensities based on a simple sum of layer-by-layer contributions neglects effects such as intensity oscillations in LEED which can occur in particular for a (0,0) peak due to island growth on a surface.[25] It seems unlikely that an analogous situation would occur for a first-order (1,0) peak, however, nor for higher order peaks. Additionally, by selecting an energy for which the peak intensity is a maximum, these types of destructive interference effects are avoided.

The graphitization technique described here is suitable both for alternative environments and larger wafer sizes. Thickness monitoring using LEED was demonstrated to accurately produce graphene films with specific thicknesses on a SiC substrate, and appears to be a suitable method for use in larger-scale graphene production.

**Acknowledgements**
This work was supported by the National Science Foundation, grant DMR-0856240.

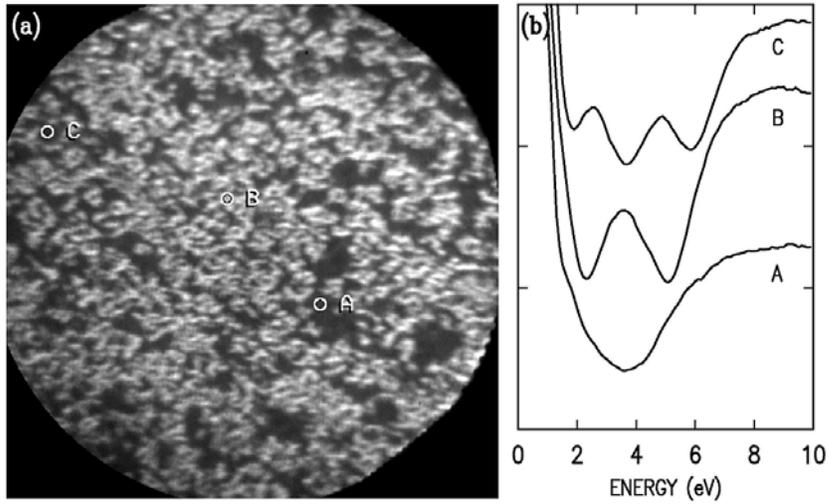

**FIG 1.** (a) LEEM image with 15 μm field-of-view, acquired with 3.7 eV incident electron energy from a graphene film on Si-face SiC prepared by annealing for 10 min at 1320°C. (b) Intensity of emitted electrons, as a function of the incident energy, acquired from the locations labeled A – C in the image. The curves have been shifted on the vertical axis, for clarity.

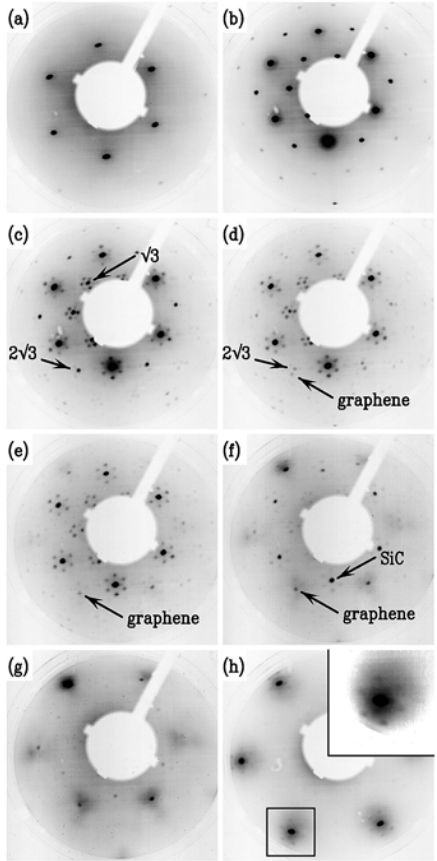

**FIG 2.** LEED patterns shown in reverse contrast, acquired from Si-face SiC after the following consecutive annealing steps: (a) bare SiC, (b) 1 min at 900°C, (c) 2 min at 1200°C, (d) 5 min at 1250°C, (e) 20 min at 1350°C, (f) 5 min at 1400°C, (g) 5 min at 1450°C, (h) no further annealing. All patterns were acquired at 100 eV electron energy except for (h) which is at 70 eV. The inset in (h) shows an expanded view of the graphene spot indicated by the square. The primary (1,0) SiC and graphene spots are indicated, as are the SiC √3 and 2√3 spots.

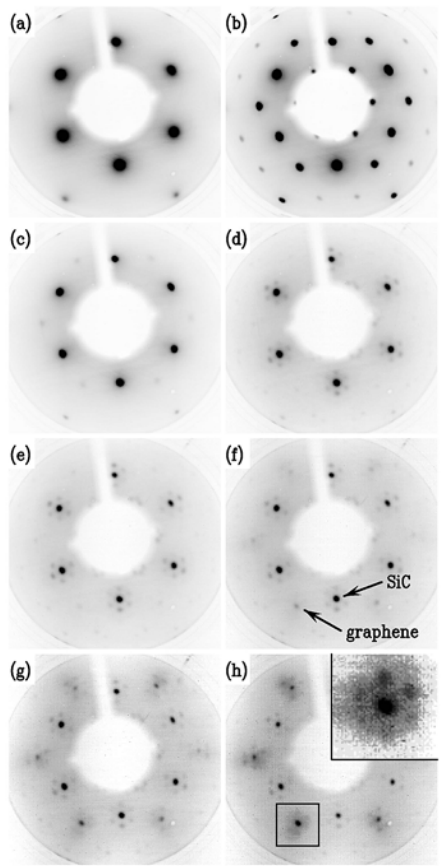

**FIG 3.** LEED patterns shown in reverse contrast, acquired from Si-face SiC after the following consecutive annealing steps: (a) base SiC, (b) after H-etching, (c) 5 min at 1160°C, (d) 5 min at 1240°C, (e) 5 min at 1290°C, (f) 5 min at 1320°C, (g) 5 min at 1360°C, (h) 5 min at 1400°C. All patterns were acquired at 100 eV electron energy except for the inset of (h), which shows an expanded view of the graphene spot indicated by the square, acquired at 133 eV. The primary (1,0) SiC and graphene spots are indicated.

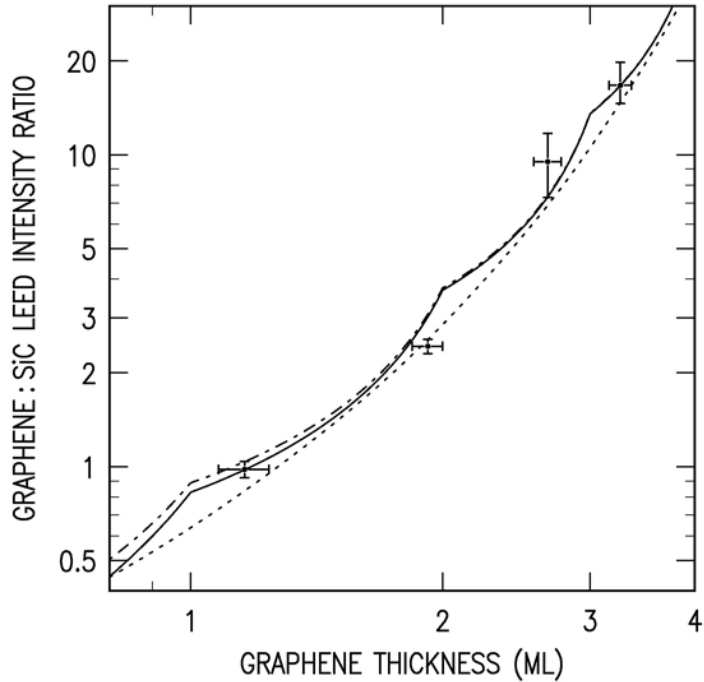

**FIG 4.** Measured values (data points) for the ratio of the graphene to SiC primary spot LEED intensity, plotted against the graphene thickness, shown on a log-log graph. Lines show results from various models: solid line – layer by layer graphene formation ($\beta = 0.29$, $I_1/I_0 = 0.24$, no buffer layer), dot-dashed line – layer by layer graphene formation ($\beta = 0.29$, $I_1/I_0 = 0.24$, with buffer layer), and dotted line – Gaussian distribution of graphene thicknesses ($\beta = 0.29$, $I_1/I_0 = 0.24$, $\sigma = 0.4$, with buffer layer).